\documentclass[11pt]{article}
\usepackage{amsmath,amssymb,color,epsf}
\usepackage{amssymb,slashed,latexsym,amsmath,multirow,color,dsfont,pifont}
\usepackage{graphics}
\usepackage{wasysym}

\usepackage{marvosym}
\usepackage{cite}
\makeatletter
\@addtoreset{equation}{section}
\makeatother

\usepackage{amsfonts}

\usepackage{float}

\usepackage{soul}

\usepackage{amsfonts}

\usepackage[font=footnotesize,labelsep=newline,labelfont=sc,justification=centering,position=top]{caption}

\newcommand{\bea}{\setlength\arraycolsep{2pt} \begin{eqnarray}}
\newcommand{\eea}{\end{eqnarray}}
\newcommand{\nn}{\nonumber}
\newcommand{\cmark}{\ding{51}}
\newcommand{\xmark}{\ding{55}}

\usepackage{hyperref}

\newcommand{\ft}[2]{{\textstyle\frac{#1}{#2}}}
\def\rmi{{\rm i}}
\newsavebox{\uuunit}
\sbox{\uuunit}
    {\setlength{\unitlength}{0.825em}
     \begin{picture}(0.6,0.7)
        \thinlines
        \put(0,0){\line(1,0){0.5}}
        \put(0.15,0){\line(0,1){0.7}}
        \put(0.35,0){\line(0,1){0.8}}
       \multiput(0.3,0.8)(-0.04,-0.02){12}{\rule{0.5pt}{0.5pt}}
     \end {picture}}

\def\be{\begin{equation}}
\def\ee{\end{equation}}
\def\ba{\begin{array}}
\def\ea{\end{array}}
\def\bea{\begin{eqnarray}}
\def\eea{\end{eqnarray}}
\def\bd{\begin{displaymath}}
\def\ed{\end{displaymath}}

\def\nn{\nonumber}

\def\a{\alpha}
\def\b{\beta}
\def\g{\gamma}

\def\d{\delta}

\def\e{\epsilon}
\def\ve{\varepsilon}

\def\vf{\varphi}

\def\p{\psi}

\def\l{\lambda}
\def\L{\Lambda}
\def\m{\mu}
\def\n{\nu}
\def\r{\rho}
\def\s{\sigma}

\def\o{\omega}

\def\x{\xi}

\def\nn{\nonumber}
\def\cD{\mathcal{D}}
\def\cN{\mathcal{N}}

\def\cM{\mathcal{M}}
\def\cJ{\mathcal{J}}

\begin{document}

\begin{titlepage}

\begin{center}

\hfill UG-15-67 \\

\vskip 1.5cm

{\Large \bf Supersymmetric Backgrounds and Black Holes in $\cN = (1,1)$ Cosmological New Massive Supergravity}
\vskip 1cm

{\bf G{\"{o}}khan Alka\c{c}\,$^1$, Luca Basanisi\,$^1$, Eric A.~Bergshoeff\,$^1$, \\[0.5ex] Deniz Olgu Devecio\u glu$^2$
and Mehmet Ozkan\,$^1$ } \\

\vskip 25pt

{\em $^1$ \hskip -.1truecm Van Swinderen Institute for Particle Physics and Gravity,  \\
University of Groningen, Nijenborgh 4, 9747 AG Groningen, The Netherlands \vskip 5pt }

{email: {\tt g.alkac@rug.nl, l.basanisi@rug.nl, e.a.bergshoeff@rug.nl, m.ozkan@rug.nl}} \\

\vskip 15pt

{\em $^2$ \hskip -.1truecm  Department of Physics, Faculty of Arts and Sciences,\\
Middle East Technical University, 06800, Ankara, Turkey \vskip 5pt }

{email: {\tt dedeveci@metu.edu.tr}} \\

\vskip 15pt

\end{center}

\vskip 0.5cm

\begin{center} {\bf ABSTRACT}\\[3ex]
\end{center}

Using an off-shell Killing spinor analysis we perform a systematic investigation of the supersymmetric background and black hole solutions of the  ${\cal N}=(1,1)$ Cosmological New Massive Gravity model. The solutions with a null
Killing vector are the same pp-wave solutions that one finds in the ${\cal N}=1$ model but we find new solutions with a time-like Killing vector that are absent in the ${\cal N}=1$ case. An example of such a solution is a Lifshitz spacetime. We also consider the supersymmetry properties of the so-called rotating hairy BTZ black holes  and logarithmic black holes in an $AdS_3$ background. Furthermore, we show that under certain assumptions there is no supersymmetric Lifshitz black hole solution.

\end{titlepage}

\newpage

\setcounter{page}{1}

\tableofcontents

\newpage

%%%%%%%%%%%%%%%%%%%%%%%%%%%%%%%%%%%%
\section{Introduction}
%%%%%%%%%%%%%%%%%%%%%%%%%%%%%%%%%%%%

Supergravity in three dimensions has a  long history. Especially in its configuration with minimal, $\cN=1$, supersymmetry the theory has been established
a long time ago both in on-shell and off-shell formulations as well as in the superconformal framework \cite{Gates:1983nr, Siegel:1979fr, Brown:1979ma, Uematsu:1984zy, Uematsu:1986de}. Despite of all the achievements in the supersymmetric constructions of the theory, the three-dimensional Poincar\'e (super)gravity by itself is of not much physical interest as the field equations of the theory imply that the spacetime curvature is zero, hence no physical degrees of freedom propagate.

The Poincar\'e theory can be supplemented with a parity-breaking Lorentz-Chern-Simons term. This combination is known as Topologically Massive Gravity (TMG)~\cite{Deser:1981wh}, and leads to a non-trivial dynamics of the gravitational field describing a helicity $+2$ or $-2$ state. The $\cN = 1$ supersymmetric completion of TMG was constructed in~\cite{DeserKay,Deser}, and the supersymmetric background and black hole solutions of this supersymmetric theory were studied in~\cite{Gibbons:2008vi}. In a subsequent development, a parity-preserving higher derivative extension of  three-dimensional gravity, known as  New Massive Gravity (NMG), was constructed~\cite{Bergshoeff:2009hq}. Similar to TMG, the NMG theory also provides dynamics to the three-dimensional gravity theory corresponding in this case to two states of helicity $+2$ and $-2$. The supersymmetric background configurations of both $\cN = 1$ TMG and $\cN =1$  NMG
 are severely restricted due to the spinor structure of the $\cN =1$ supersymmetry, which allows only planar-wave type solutions with a null Killing vector as well as maximally supersymmetric $AdS_3$ and Minkowski backgrounds \cite{Andringa:2009yc,Bergshoeff:2010mf}.

In a recent paper some of us have formulated all four-derivative extension of the three-dimensional~$\cN = (1,1)$ off-shell cosmological Poincar\'e supergravity theory~\cite{Alkac:2014hwa}. For a discussion of this construction (and more) from a superspace point of view, see
\cite{Kuzenko:2015jda}. Extending the~$\cN = 1$ theory with more supersymmetry cannot affect the dynamics of the Poincar\'e supergravity theory. It merely extends the size of the $\cN =1$ Poincar\'e multiplet, consisting of a dreibein~$e_\m{}^a$, a gravitino $\p_\m$ and a scalar $A$ with an additional gravitino, an auxiliary vector~$V_\m$ and a pseudo-scalar~$B$. As we will show in this paper, the merit of the~$\cN = (1,1)$ theory is that the spinors of the theory are Dirac instead of Majorana spinors, which allows a larger variety of supersymmetric background solutions than in the $\cN=1$ case~\cite{Deger:2013yla, Kuzenko:2013uya}.

The main aim of this paper is to study the supersymmetric backgrounds as well as black hole solutions of the~$\cN = (1,1)$ cosmological NMG, or shortly $\cN=1$ CNMG, theory~\cite{Alkac:2014hwa} using the off-shell Killing spinor analysis. The power of the off-shell analysis is reflected by the fact that, once the conditions on the possible field configurations are obtained by using the off-shell supersymmetry transformation rules, one can use them to study the solutions of any model which respects the same set of transformation rules. This might include higher derivative corrections and/or matter couplings.

We begin our study in section~\ref{CNMG} with a brief review of~$\cN = (1,1)$ CNMG and its off-shell transformation rules. As a typical property of off-shell supergravity theories, the auxiliary fields of the theory start to propagate \footnote{There are exceptional cases where the auxiliary fields do not propagate such as in $\cN = 1$ CNMG \cite{Andringa:2009yc}.} when the Poincar\'e supergravity is extended with higher-order curvature terms. Assuming that the supersymmetric theory admits at least one Killing spinor, we  present the Killing spinor equation and its integrability condition. We then review the implications of
the existence of an off-shell Killing spinor as presented in~\cite{Deger:2013yla}. The existence of such a spinor imposes  numerous algebraic as well as differential identities on the metric, the vector $V_\m$ and the scalars $A$ and $B$ of the theory. These identities are the backbone of our analysis that we present in the remainder of the paper. The Killing spinor equation in particular  implies that the background solutions can be put into two categories depending on whether the Killing vector that is formed out of the Killing spinors is {\sl null} or {\sl timelike}.

In section~\ref{Null}, we investigate the solutions that admit a null Killing vector. In this case, the analysis for finding supersymmetric solutions  simply reduces to the one corresponding to the~$\cN = 1$ theory \cite{Andringa:2009yc} as the vector~$V_\m$ and the pseudo-scalar~$B$ are set to zero due to the algebraic and differential constraints, which are consequences of the existence of a Killing spinor. The solution are, therefore, of the pp-wave type, like in the $\cN=1$ case.

In the timelike case, which we present in section \ref{Timelike}, the supersymmetric solutions are categorized according to the values of the components $V_a$ of the vector in a flat basis. We find that the $\cN = (1,1)$ CNMG theory allows all solutions of $\cN = (1,1)$ TMG~\cite{Deger:2013yla} with shifted parameters. Furthermore, we find additional~$AdS_2 \times \mathbb{R}$ and Lifshitz solutions. These extra solutions are possible because of the fact that in~$\cN = (1,1)$ TMG, the vector equation gives rise to a second order algebraic equation for the components~$V_a$, whereas in the case of $\cN = (1,1)$ CNMG, the resulting equation is cubic, allowing more solutions.

In section~\ref{BlackHoles}, we investigate the supersymmetric black holes with AdS$_3$ and Lifshitz backgrounds.  We first show that the rotating hairy BTZ black hole of \cite{Giribet:2009qz}, which is a generalization of the well-known BTZ black hole~\cite{Banados:1992wn} obtained by introducing a \emph{gravitational hair} parameter, and the \emph{logarithmic black hole} of~\cite{Clement:2009ka} are solutions of the~$\mathcal{N}=(1,1)$ CNMG theory for the extremal cases. This is also true for the rotating BTZ black hole which can be  obtained by setting the hair parameter $b$ to zero.  Given the Lifshitz solution, we then analyze whether we can find an extremal Lifshitz black hole. As the theory is ungauged, one can hope to saturate the BPS bound with the \emph{massive vector hair} $V_\m$. As opposed to the pseudo-supersymmetry analysis of the Einstein-Weyl theory in four dimensional $\cN = 1$ supergravity \cite{Lu:2012am}, this is not the case in $\cN = (1,1)$ CNMG. Furthermore, we will show that a simple rotating black hole ansatz fails to satisfy the Killing spinor equation and the field equations of the theory simultaneously.

Finally, in section \ref{Conclusions}, we present our conclusions and discuss further directions.

%%%%%%%%%%%%%%%%%%%%%%%%%%%%%%%%%%%%
\section{$\cN = (1,1)$ Cosmological New Massive Supergravity}{\label{CNMG}}
%%%%%%%%%%%%%%%%%%%%%%%%%%%%%%%%%%%%

The field content of the $\cN = (1,1)$ supergravity theory consists of the dreibein $e_\m{}^a$, the gravitino~$\p_\m$, a complex scalar $S$, and a vector $V_\m$. The model we shall study is a particular combination of supersymmetric invariants up to dimension four \cite{Alkac:2014hwa} that leads to a model that, when expanded around a supersymmetric~$AdS_3$ vacuum, propagates only helicity $\pm\, 2$ and $\pm\, 3/2$ states with AdS energies that respect perturbative unitarity.
This model is called cosmological New Massive Gravity~(CNMG). Here we focus on the bosonic part of the supersymmetric CNMG Lagrangian which is given by
\bea
e^{-1} {\mathcal L}_{\rm CNMG} &=&  \s (R+2V^2 -2|S|^2 ) + 4 M A
\nn\\[.1truecm]
&& +\frac{1}{m^2} \Big[ R_{\mu\nu} R^{\mu\nu} -\ft38 R^2   - R_{\mu\nu} V^\mu V^\nu -  F_{\mu\nu} F^{\mu\nu} +\ft14 R(V^2 -B^2)
\nn\\[.1truecm]
&& \qquad+ \ft16 |S|^2(A^2-4B^2) -\ft12 V^2 (3A^2+4B^2) - 2V^\mu  B\partial_\mu A \Big]\,,
\label{NMG11}
\eea
where $(\s,M, m^2)$  are arbitrary real constants and we have defined $S=A+\rmi B$. The action corresponding to this Lagrangian is invariant under the following off-shell supersymmetry transformation rules \footnote{In this paper, we follow the conventions of \cite{Deger:2013yla}, with the only difference being that the $S$ we are using here is replaced by $S \rightarrow -Z$.  }
\bea
\delta e_\mu{}^{a}  &= & \ft{1}{2}\bar{\epsilon}\gamma^{a}\psi_{\mu}+h.c.\,,
\nn\\[.1truecm]
\delta\psi_\mu  &= & D_{\mu}(\widehat\omega)\,\epsilon-\ft{1}{2} \rmi V_{\nu}\,\gamma^{\nu}\gamma_{\mu}\,\epsilon
-\ft12 S\gamma_\mu \epsilon^{\star}\,,
\nn\\[.1truecm]
\delta V_\mu  &= & \ft{1}{8} \rmi \bar{\epsilon}\,\gamma^{\nu\rho}\gamma_{\mu}\left(\psi_{\nu\rho}-{\rmi} V_{\sigma}\gamma^{\sigma}\gamma_{\nu}\,\psi_{\rho}-S\gamma_\n \psi_{\rho}^{\star} \right)+h.c.\,,
\nn\\[.1truecm]
\delta S  & = & -\ft14 \tilde{\epsilon}\,\gamma^{\mu\nu}\left(\psi_{\mu\nu}
-\rmi V_{\s}\,\gamma^{\s}\gamma_{\mu}\psi_{\nu}-S\gamma_{\mu} \psi_{\nu}^{\star}\right)\ ,
\label{TransformationRules}
\eea
where $ \tilde{\epsilon} = \overline{\e^\star}\,,\,\,{\widehat\omega}$ is the super-covariant spin-connection and
\be
D_{\mu}(\widehat\omega)\epsilon=(\partial_{\mu}+\ft14\widehat\omega_{\mu}{}^{ab}\,\gamma_{ab})\epsilon\ ,
\qquad\psi_{\mu\nu}=2D_{[\mu}(\widehat\omega)\psi_{\nu]}\ .
\ee
The transformation rules (\ref{TransformationRules}) are off-shell as the algebra closes on these fields without imposing the field equations
 corresponding to the Lagrangian (\ref{NMG11}).

 In order to determine the supersymmetric backgrounds allowed by  a model with the transformation rules (\ref{TransformationRules}), one considers the Killing spinor equation
\bea
\cD_\m \e = \partial_\m \e +\ft14\widehat\omega_{\mu}{}^{ab}\,\gamma_{ab}\epsilon  -\ft{1}{2} \rmi V_{\nu}\,\gamma^{\nu}\gamma_{\mu}\,\epsilon
-\ft12 S\gamma_\mu \epsilon^{\star} = 0 \,.
\label{KillingSpinor}
\eea
Any Killing spinor $\e$ satisfying this equation must also satisfy  the integrability condition
\bea
[\cD_\m, \cD_\n] \e &=& \frac{1}{4}   \Big( R_{\m\n}{}^{\r\s} + 2 \d_\m^\r \d_\n^\s (A^2 + B^2)  + 2 \d_\m^\r \d_\n^\s V^2 - 4 \rmi \d_{[\n}^\s \nabla_{\m]} V^\r - 4 \d_{[\n}^\s V_{\m]} V^\r \Big) \g_{\r\s} \e  \nn\\
&& - \delta_{[\n}^\s \Big(  \partial_{\m]}A +  B V_{\m]}  \Big) \g_\s \e^* - \rmi  \delta_{[\n}^\s  \Big( \partial_{\m]} B - A V_{\m]} \Big) \g_\s \e^* - \frac12 \rmi F_{\m\n} \e \nn\\
&& + \rmi \e_{\m\n\r} V^\r (A + \rmi B) \e^* =0\,.
\label{integrability}
\eea
Considering the field equations for $A, B, V_\m$ and $g_{\m\n}$,
\begin{eqnarray}
0 &=& 4 M - 4\s A + \frac{1}{m^2} \left[ \frac{2}{3} A^3 - B^2 A - 3V^2 A + 2 \left( \nabla \cdot V \right) B + 2 V^{\mu} \partial_{\mu} B \right]  \,, \nn \\
0 &=& 4\s B + \frac{1}{m^2} \left[ \frac{1}{2} R B + A^2 B + \frac{8}{3} B^3 + 4 V^2 B + 2 V^{\mu} \partial_{\mu} A \right] \,, \nn \\
0 &=&  4\s V_{\mu} - \frac{1}{m^2} \left[ 2 R_{\mu\nu} V^{\nu} + 4 \nabla_{\nu} F_{\mu}{}^{\nu} + V_{\mu} \left( 3A^2 + 4B^2 - \frac{R}{2} \right) + 2 B\partial_{\mu} A \right] \nn\,,\\
0 &=&  \s \Big( R_{\mu\nu} +2  V_{\mu}V_{\nu} - \frac{1}{2} g_{\mu\nu} [ R + 2V^2 - 2(A^2 + B^2) ] \Big) -2 g_{\mu\nu} M A   \qquad \qquad \nn \\
&&  + \frac{1}{m^2} \Bigg[ \Box R_{\mu\nu} - \frac{1}{4} \nabla_{\mu}\nabla_{\nu} R + \frac{9}{4} R R_{\mu\nu} - 4 R^{\rho}_{\mu} R_{\nu \rho} - 2 F_{\mu}{}^{\rho} F_{\nu \rho}  \qquad \nn \\
&& + \frac{1}{4} R V_{\mu}V_{\nu} - 2 R_{(\mu}^{\rho} V_{\nu)} V_{\rho} - \frac{1}{2} \Box (V_{\mu}V_{\nu}) + \nabla_{\rho} \nabla_{(\mu} (V_{\nu)} V^{\rho}) \qquad \nn \\
&& + \frac{1}{4} R_{\mu\nu} (V^2 - B^2) - \frac{1}{4} \nabla_{\mu} \nabla_{\nu} (V^2 - B^2) - \frac{1}{2} V_{\mu}V_{\nu} (3A^2 + 4B^2) \qquad \nn \\
&&  - 2B V_{(\mu} \partial_{\nu)} A - \frac{1}{2} g_{\mu\nu} \Big( \frac{13}{8} R^2 + \frac{1}{2} \Box R - 3 R_{\rho \sigma}^2 - R_{\rho \sigma} V^{\rho} V^{\sigma}  \qquad \nn \\
&&   + \nabla_{\rho} \nabla_{\sigma} (V^{\rho} V^{\sigma}) - F_{\rho \sigma}^2 + \frac{1}{4} R (V^2 - B^2) - \frac{1}{2} \Box (V^2 - B^2) \qquad \nn \\
&&   + \frac{1}{6} (A^2 + B^2) (A^2 - 4B^2) - \frac{1}{2} V^2 (3A^2 + 4B^2) - 2 B V^{\rho}\partial_{\rho} A \Big) \Bigg] \,,
\label{fieldequations}
\end{eqnarray}
it can be seen that for  cosmological Poincar\'e supergravity, i.e.~$m \rightarrow \infty$, $A, B$ and $V_\m$ can be eliminated algebraically. In this case, the integrability condition (\ref{integrability}) reduces to
\bea
\Big( R_{\m\n}{}^{\r\s} + 2 \d_\m^\r \d_\n^\s M^2  \Big) \g_{\r\s} \e = 0 \,,
\eea
which implies that the maximally supersymmetric background is either Minkowski with $M=0$, or $AdS_3$ with radius $1/M^2$.  More solutions, with less supersymmetry, can be obtained by imposing projection conditions on $\e$. Note that even with the higher derivative contributions, the maximally supersymmetric solution is still given by the same background solution with a shifted value of the cosmological constant. The reason for this is that  the expectation value of $A$  receives a contribution from the higher derivative corrections whereas~$B$ and~$V_\m$ do not and, therefore, can still be set to zero.

In the case of cosmological Poincar\'e supergravity, the auxiliary fields can be eliminated from the theory, resulting in an on-shell supergravity theory with the field content $(e_\m{}^a, \p_\m)$. However, with the higher derivative contributions added, the massive vector and the real scalars become dynamical and hence cannot be solved algebraically. These `auxiliary' fields play a crucial role in determining the supersymmetric backgrounds allowed by the CNMG Lagrangian (\ref{NMG11}).

Now that we have clarified the maximally supersymmetric backgrounds, let us proceed to the case where we have at least one unbroken supersymmetry. In order to do so, we will briefly review the implications of an off-shell Killing spinor following the discussion of  \cite{Deger:2013yla}. From the symmetries of the gamma matrices,  one finds the following identities for a commuting Killing spinor $\e$
\bea
\bar{\e} \e^\star = \tilde{\e} \e = 0 \,.
\eea
Thus, non-vanishing spinor bilinears can be defined as follows
\bea
\bar{\e}\e  = - \tilde{\e} \e^\star \equiv \rmi f \,, \qquad \bar{\e}\g_\m \e = \tilde{\e} \g_\m \e^\star \equiv K_\m \,, \qquad \bar{\e} \g_\m \e^\star &\equiv& L_\m = S_\m + \rmi T_\m \,,
\label{SpinorBilinears}
\eea
where $f$ is a real function and $K_\m\, (L_\mu)$ is a real (complex) vector.
 Using the Fierz identities for commuting spinors, one can show that
\bea
&& K_\m K^\m = - f^2\,, \qquad K_\m \g^\m \e = \rmi f \e \,.
\label{KST}
\eea
The first equation implies that the vector is either null or timelike. Using the Killing spinor equation (\ref{KillingSpinor}) one finds that
\bea
\nabla_{(\m} K_{\n)} = 0 \,,
\label{KillingVector}
\eea
showing that $K_\m$ is a Killing vector. Finally, we may derive the following differential identities following from the Killing spinor equation \eqref{KillingSpinor}
\bea
\partial_{[\mu} K_{\nu]} &=&
\e_{\mu\nu\rho}\Big(
-f V^\rho - \frac12\left(S L^\rho+S^\star (L^\star)^\rho\right)\Big)\;
\label{KILdifDmuK} \,,\\[.1truecm]
\partial_{\mu} f &=&
- \e_{\mu\nu\rho} V^\nu K^\rho - \frac{1}2 \rmi \left(S L_\mu-S^\star L^\star_\mu \right)
\;.
\label{KILdifdmuf}
\eea
We refer to \cite{Deger:2013yla} for readers interested in the derivation of these Killing spinor identities and of other implications of the existence of a Killing spinor.

\section{The Null Killing Vector}{\label{Null}}
We first consider the case that the function $f$ introduced in eq.~(\ref{SpinorBilinears}) is zero everywhere, i.e.~$f=0$.
 This implies that $K_\mu$ is a  null Killing vector. The case that $f\ne 0$ will be discussed in the next section.  In our conventions, a Majorana spinor field has all real components. This being said, the first spinor bilinear equation in (\ref{SpinorBilinears}) leads to a Dirac spinor $\e$ that is proportional to a real spinor $\e_0$ up to a  phase factor
characterized by an angle $\theta$ \cite{Deger:2013yla},
\bea
\e = e^{-\rmi \ft{\theta}{2}} \, \e_0 \,.
\label{NullEpsilon}
\eea
The above equation implies that  $L_\m = e^{\rmi \theta} K_\m$. Taking this into account,
the differential equation (\ref{KILdifDmuK}) reads
\bea
\partial_{[\mu} K_{\nu]} =
-    \rm{Re}(S e^{\rmi \theta})\, \e_{\mu\nu\rho}\, K^\rho \; \,.
\eea
Contracting this equation with $K^{\mu}$ we find that
\bea
K^\m \, \nabla_\m \, K_\n = 0 \,.
\label{KNK}
\eea
The same equation also implies that  $K \wedge dK = 0$, i.e.~$K$ is hypersurface orthogonal. Thus, there exist functions $u$ and $P$ of the
three-dimensional spacetime coordinates such that
\bea
K_\m \, dx^\m = P du \,.
\label{KmPu}
\eea
Eq.~(\ref{KNK}) implies that that $K$ is tangent to affinely parameterized geodesics in the surface of constant $P$. One can, then, choose coordinates $(u,v,x)$ such that $v$ is an affine parameter along these geodesics, i.e.
\bea
K^\m \, \partial_\m = \frac{\partial}{\partial v} \,.
\eea
By virtue of our choice for $K_\m$ the metric components further simplify to
\bea
g_{uv} = P(u,x), \qquad g_{vv} = g_{xv} = 0 \,,
\eea
where $P = P(u,x)$ since we demand the null direction to be along the $v$ direction. All these choices yield a metric of the following generic form
\bea
ds^2 = h_{ij}(x,u)\, dx^i \, dx^j + 2 P(x,u)\, du\, dv \,,
\eea
where $x^i = (x,u)$. Without loss of generality, this metric can be cast in the following form by a coordinate transformation \cite{Gibbons:2008vi,Andringa:2009yc}
\bea
ds^2 = dx^2 + 2 P (x,u)\, du\, dv + Q(x,u)\, du^2 \,,
\label{NullMetric1}
\eea
with $\sqrt{|g|} = P$. With these results in hand, the auxiliary fields of the theory should satisfy the following constraints \cite{Deger:2013yla}
\bea
V_\m &=& - \frac{1}{2} \partial_\m\, \theta (x,u) \,, \nn\\
S e^{\rmi \theta} + S^\star e^{-\rmi \theta} &=& \partial_x \log P(x,u) \,.
\label{NullConditions}
\eea

In the next subsection   we will investigate the solutions of CNMG under the assumption that $f=0$.

\subsection{The General Solution}

 To find the general solution with $f=0$  we set $S$ to be a constant, to be precise we set $A = - \frac{1}{l}$ and $ B = 0$. Using (\ref{KILdifdmuf}) we obtain
\bea
\e_{\m\n\r}\, V^\n K^\r = - \frac1l K_\m \sin\theta(u,x) \,.
\eea
The $u$ component of this equation reads
\bea
\frac1l K_u \sin\theta(u,x) = P(u,x) V_x \,,
\eea
where we have used that $\ve_{xuv} = 1$. Provided that the function $P(u,x)$ is nowhere vanishing,
it is straightforward to integrate the first (vector) equation in (\ref{NullConditions}) and obtain
\bea
\theta(u,x) = \arctan \Big(  \frac{2\,c(u)\, e^{-2x/l}}{1 - c^2 (u)\,e^{-4x/l} } \Big)\,,
\label{theta1}
\eea
for arbitrary $c(u)$. From the second (scalar) equation in (\ref{NullConditions}) we deduce  that
\bea\label{eq1}
- \frac2l \cos\theta(u,x) = \partial_x \log P(u,x) \,,
\eea
which, upon using eq.~(\ref{theta1}), yields
\bea\label{eq2}
P(x,u) = P(u) [ \,e^{2x/l} +  e^{-2x/l} c^2 (u) ] \,,
\eea
where $P(u)$ is an arbitrary  function of $u$. We may set $P(u)$ to unity without loss of generality \cite{Andringa:2009yc}. Using eqs.~\eqref{eq1} and \eqref{eq2} in the vector field equation \eqref{theta1}, we deduce that~$c(u) = 0$ and~$\theta(u,x) = n \pi$. In order to fix $n$ we use to the trace of the gravity equation and find that~$\theta(u,x) = \pi$.

We thus find that the metric (\ref{NullMetric1}) takes the following final form
\bea
ds^2 = dx^2 + 2\, e^{2x / l}\, du\, dv + Q(x,u)\, du^2 \,.
\label{ppwavemetric}
\eea
This is the general form of a pp-wave metric. Taking the limit $l \rightarrow \infty$ gives rise to the pp-wave in a Minkowski background. Setting $l = 1$ and substituting $A = -1$, $B = 0$, $V_x = V_u = V_v = 0$  into the metric field equation, we find that $Q(x,u)$ satisfies the following differential equation
\bea
(2+ 4 \s m^2)\, Q^\prime - ( 9 + 2 \s m^2)\, Q^{\prime\prime} + 8 Q^{\prime \prime \prime} - 2 Q^{\prime\prime\prime\prime} = 0 \,,
\label{DiffEqQ}
\eea
where the prime denotes a derivative with respect to $x$. The most general solution of this differential equation
is given by
\bea\label{Qsolution}
Q(x,u) = e^{(1- \sqrt{ \ft12 - \s m^2  })x}\, C_1(u) +  e^{(1 + \sqrt{ \ft12 - \s m^2  })x}\, C_2(u) + e^{2x}\, C_3(u) + C_4 (u) \,,
\eea
where the functions $C_i (u)\,,i=1,\cdots,4,$ are arbitrary functions of $u$. We note that this expression for $Q(x,u)$  matches with that of \cite{ Bergshoeff:2010mf,AyonBeato:2009yq}. It differs, however, with the expression given in \cite{Andringa:2009yc}. This is  due to the fact that the off-diagonal coupling of gravity to the scalar~$A$ was included in the supersymmetric New Massive Gravity model studied in \cite{Andringa:2009yc}, whereas such a term is absent in our case, see  eq.~(\ref{NMG11}).

The solution for $Q(x,u)$ given in \eqref{Qsolution} has a redundancy \cite{Gibbons:2008vi}. To make this redundancy manifest  we consider the following coordinate transformation
\bea
x = \widetilde{x} - \ft12 \log a^\prime\,, \qquad u = a(\widetilde{u})\,, \qquad v = \widetilde{v} - \ft14 e^{-2\widetilde{{x}}}\, \frac{a^{\prime\prime}}{a^\prime} + b(\widetilde{u}) \,,
\eea
where $a(\widetilde{u})$ and $b(\widetilde{u})$ are arbitrary functions of $\widetilde{u}$ and the prime denotes a derivative with respect to $\widetilde{u}$. By choosing the function $a(\widetilde{u})$ and $b(\widetilde{u})$ such that
the differential equations \bea
\Big( \frac{a^{\prime\prime}}{a^\prime} \Big)^\prime - \frac12  \Big( \frac{a^{\prime\prime}}{a^\prime} \Big)^2
- 2 (a^\prime)^2\, \widetilde{C}_4 (\widetilde{u}) = 0 \,, \qquad b^\prime + \frac12 a^\prime \, \widetilde{C}_3 (\widetilde{u}) = 0\,,
\eea
are satisfied  the functions $\widetilde{C}_3$ and $\widetilde{C}_4$ can be set to zero. This implies that, without loss of generality, we may set $C_3 = C_4 = 0$. In addition to this, we get
\bea
\widetilde{C}_1(\widetilde{u}) = C_1 ( a(\widetilde{u}))\, [a^\prime (\widetilde{u})]^{\ft12 (3 + \sqrt{ \ft12 - \s m^2  })} \,, \quad \widetilde{C}_2(\widetilde{u}) = C_2 ( a(\widetilde{u}))\, [a^\prime (\widetilde{u})]^{\ft12 (3 - \sqrt{ \ft12 - \s m^2  })} \,.
\eea

There are two special values of parameters which must be handled separately. These are the cases  $\s m^2 = \pm \ft12$. The reason for this is that  for the $\s m^2 =  \ft12$ case the function $C_1$ degenerates with $C_2$ whereas for the $\s m^2 = - \ft12$ case the function $C_1$ degenerates with $C_4$ while the function $C_2$ degenerates with $C_3$. Therefore, we solve the field equation (\ref{DiffEqQ}) for these special cases, and display the solutions $Q(x,u)$ for these special values of the  parameters  explicitly:
\bea
\s m^2 =\ft12 &:& \qquad Q(x,u) = e^x \, D_1 (u) + x\, e^x \, D_2 (u) + e^{2x}\, D_3(u) + D_4 (u)      \,,\nn\\
\s m^2 = - \ft12 &:& \qquad Q(x,u) =  x\, e^{2x}\, D_1 (u) + x\, D_2 (u) + e^{2x}\, D_3(u) + D_4 (u)     \,.
\eea
Here $D_i(u)\,,\, i=1,\dots,4$, are arbitrary functions of $u$. Setting $D_3 = D_4 = 0$, we are led to the following cases:
\bea
\s m^2 \neq \pm\ft12 &:&  ds^2 = dx^2 + 2\, e^{2x}\, du\, dv + \Big( e^{(1- \sqrt{ \ft12 - \s m^2  })x}\, D_1(u) +  e^{(1 + \sqrt{ \ft12 - \s m^2  })x}\, D_2(u) \Big)\, du^2 \,,\nn\\[.1truecm]
\s m^2 =\ft12 &:& ds^2 = dx^2 + 2\, e^{2x}\, du\, dv + \Big( e^x\, D_1 (u) + x\, e^x\, D_2 (u) \Big)\, du^2 \,,\nn\\[.1truecm]
\s m^2 =-\ft12 &:& ds^2 = dx^2 + 2\, e^{2x}\, du\, dv + \Big( x\, e^{2x}\, D_1 (u) + x\, D_2 (u) \Big) \,du^2 \,.
\label{ppwave}
\eea
These pp-wave solutions coincide with the solutions of $\cN = 1$ CNMG \cite{Bergshoeff:2010mf}. Having found the most general solutions for the null case, we will  continue in the next subsection with determining the amount of supersymmetry that these solutions preserve by working out the Killing spinor equation (\ref{KillingSpinor}).

\subsection{Killing Spinor Analysis}
In order to construct the Killing spinors for the pp-wave metric (\ref{ppwavemetric}) we introduce the following orthonormal frame   \cite{Gibbons:2008vi}
\bea
e^0 = e^{\ft{2x}{l} - \b}\, dv, \qquad e^1 = e^\b du + e^{\ft{2x}{l} - \b}\, dv, \qquad e^2 = dx \,,
\eea
where $Q(u,x) = e^{2\b(u,x)}$. It follows that the components of the spin-connection are given by
\bea
\o_{01} &=& - \dot{\b}\, du - \Big(\b^\prime - \frac1l \Big) dx \,,\nn\\
\o_{02} &=& - \Big( \b^\prime - \frac1l \Big) e^\b \, du - \frac1l \, e^{\ft{2x}{l} - \b}\, dv \,,\nn\\
\o_{12} &=& \b^\prime \, e^\b \, du + \frac1l \, e^{\ft{2x}{l} - \b}\, dv \,,
\eea
where
\bea
\dot{\b} \equiv \frac{\partial \b}{\partial u} \,, \qquad \b^\prime \equiv \frac{\partial \b}{\partial x} \,.
\eea
For the null case, the Killing spinor equation (\ref{KillingSpinor}) then reads
\bea
0 = d \e + \frac14 \,\o_{ab}\, \g^{ab} \e + \frac1{2l} \, \g_a \,e^a \, \e^\star  \,.
\label{NullKS}
\eea
%Imposing the condition  $\e = -\rmi \e_0$ due to (\ref{NullEpsilon}), and by the
We make the following choice of the $\g$ matrices
\bea
\g_0 = \rmi \s_2 \,, \qquad \g_1 = \s_1 \,, \qquad \g_2 = \s_3 \,,
\eea
where $\s_i$'s are the standard Pauli matrices. With this choice the Killing spinor equation reads
\bea
0 = d \e &+&  \ft12 \Big( \dot{\b} \, \s_3 \,\e - e^\b \b^\prime (\s_1 + \rmi \s_2)\,\e + \frac1l e^\b \s_1 \,(\e +  \e^\star) \Big)  du \nn\\[.1truecm]
&-& \frac1{2l}\, e^{\ft{2x}{l} - \b} \,(\s_1 + \rmi \s_2) \,(\e -  \e^\star) \,       dv \nn\\[.1truecm]
&+& \frac12 \, \Big( \b^\prime \s_3 \,\e -  \frac1l \, \s_3 \, (\e - \e^\star)       \Big) \, dx \,.
\label{KillingSpinor3}
\eea
Decomposing a Dirac spinor into two Majorana spinors as $\e = \xi + \rmi \zeta$, i.e.
\bea
\e =
\begin{pmatrix}
\xi_1 + \rmi \zeta_1 \\
\xi_2 + \rmi \zeta_2
\end{pmatrix} \,,
\label{DiracMajorana}
\eea
we find the following equations for the components
\bea
0 &=& d \xi_1 + \frac12 \dot{\b}\, \xi_1 \, du - e^\b (\b^\prime - \frac1l)\, \xi_2 \, du + \frac12\, \xi_1\, \b^\prime \, dx \,,\nn\\[.1truecm]
0 &=& d \xi_2 + \frac1l \, e^\b\, \xi_1\, du - \frac12 \, \dot{\b} \, \xi_2 \, du - \frac12 \, \b^\prime \, \xi_2 \, dx \,,\nn\\[.1truecm]
0 &=& d \zeta_1 + \frac12 \, \dot{\b} \, \zeta_1 \, du - e^\b \, \b^\prime \, \zeta_2 \, du - \frac2l \, e^{\ft{2x}{l} - \b} \, \zeta_2 \, dv + \frac12 \, (\b^\prime - \frac2l) \, \zeta_1 \, dx \,,\nn\\[.1truecm]
0 &=& d \zeta_2 - \frac12 \, \dot{\b} \, \zeta_2 \, du - \frac12 \, (\b^\prime - \frac2l) \, \zeta_2 \, dx  \,.
\label{KSAdsPPWave}
\eea
The first two equations are uniquely solved by $\xi_1 = \xi_2 = 0$. For the last two equations, the solution for a generic function $\b(u,x)$ is given by
\begin{equation}
\zeta_1 = e^{-\ft12\b +  \ft{x}{l} }\,,\hskip 1.5truecm \zeta_2 = 0\,.
 \end{equation}
 There is an additional solution for the special case that $\b = x$. It is given by
 \begin{equation}
 \zeta_1 = (u + 2v) e^{\frac12 x}\,,\hskip 1.5truecm \zeta_2 = e^{-\frac12 x}\,.
  \end{equation}
  This solution corresponds to the first case given in eq.~(\ref{ppwave}) with $D_1 (u) = 0$ and $D_2 (u) = 1$.
   There is, however, a problem with this solution. One must  choose $\s m^2 = - \frac12$ for this solution and this conflicts with the condition imposed on this pp-wave solution when we classified the different solutions in the previous subsection. Therefore, we conclude that the pp-wave Killing spinor equation is uniquely solved by
   \begin{equation}
   \xi_1 = \xi_2 = \zeta_2 = 0\,,\hskip 1.5truecm \zeta_1 = e^{-\ft12\b +  \ft{x}{l} }\,.
    \end{equation}
     This implies that the pp-wave solutions all preserve  $1/4$ of the supersymmetries. Note that in the Minkowski limit $l \rightarrow \infty$, the equations for $\xi$ and $\zeta$ degenerate. Thus, the number of Killing spinors are the same for both $AdS$ and Minkowski pp-wave solutions.

We conclude this section by noting that when $D_1 = D_2 = 0$, the metric reduces to
\bea
ds^2 = dx^2 + 2 e^{2x/l}\, du\, dv = dx^2 + e^{2x/l} \, ( - dt^2 + d \phi^2) \,,
\label{AdSMetric}
\eea
which is the $AdS_3$ metric in a Poincar\'e patch. In this case, we have
\bea
e^0 = e^{x/l} \, dt \,, \qquad e^1 = e^{x/l}\, d\phi \,, \qquad e^2 = dx \,.
\eea
which implies that
\bea
\o_{02} = - \frac1l \, e^{x/l} \, dt \,, \qquad \o_{12} = \frac{1}{l}\, e^{x/l}\, d \phi \,.
\eea
The Killing spinor equation then turns into
\bea
d\e - \frac1{2l} \, e^{x/l} \Big( \s_1 \e - \rmi \s_2 \e^\star   \Big) dt - \frac1{2l} \, e^{x/l} \Big( \rmi \s_2 \e - \s_1 \e^\star  \Big) d\phi + \frac1{2l} \, \s_3 \, \e^\star dx = 0 \,.
\eea
Decomposing the Dirac spinor into two Majorana spinors as $\e = \xi + \rmi \zeta$, see eq.~\eqref{DiracMajorana},
the Killing spinor equation gives rise to the following equations
\bea
0 &=& d \xi_1 + \ft1{2l}\, \xi_1 dx \,,\nn\\[.1truecm]
0 &=& d \xi_2 - \ft1l \, e^{x/l} \, \xi_1 dt + \ft1l \, e^{x/l} \, \xi_1 d \phi - \ft1{2l} \, \xi_2 dx \,,\nn\\[.1truecm]
0 &=& d \zeta_1 - \ft1l \, e^{x/l} \, \zeta_2 \, dt  - \ft1l \, e^{x/l} \, \zeta_2 \, d \phi - \ft1{2l} \, \zeta_1 dx \,,\nn\\[.1truecm]
0 &=& d \zeta_2 + \ft1{2l} \,\zeta_2 \, dx \,.
\label{KSAds}
\eea
 Making use of the fact that that the $\xi$ and $\zeta$ equations are decoupled from each other, we find the following four independent solutions:
\begin{enumerate}
\item{$\xi_1 = 0, \quad \x_2 = e^{\ft{x}{2l}}, \quad \zeta_1 = \zeta_2 = 0 $,}
\item{$\xi_1 = e^{-\ft{x}{2l}}, \quad \x_2 = \ft1l\, e^{\ft{x}{2l}} (t - \phi), \quad \zeta_1 = \zeta_2 = 0 $,}
\item{$\xi_1 = \x_2 = 0, \quad \zeta_1 = e^{\ft{x}{2l}}, \quad \zeta_2 = 0 $,}
\item{$\xi_1 = \x_2 = 0, \quad \zeta_1 = \ft1l\, e^{\ft{x}{2l}} (t + \phi), \quad \zeta_2 = e^{-\ft{x}{2l}} $,}
\end{enumerate}
Therefore, the $AdS_3$ solution has a supersymmetry enhancement with four Killing spinors.

\section{The Timelike Killing Vector}{\label{Timelike}}

In this section, we shall consider the case that $f \neq 0$ and hence that $K$ is a timelike Killing vector field. Introducing a coordinate $t$ such that $K^\m \partial_\m = \partial_t$, the metric can be written as \cite{Deger:2013yla}
\be
ds^2= - e^{2\varphi(x,y)} \left(dt+ B_{\alpha}(x,y)\, dx^{\alpha} \right)^2 + e^{2\l(x,y)} (dx^2+dy^2)\ ,
\label{metric}
\ee
where $\lambda(x,y)$ and $\vf(x,y)$ are arbitrary functions and $B_\a \, (\alpha=x,y)$ is a vector with two components. The Dreibein corresponding to this metric is naturally written as
\begin{equation}\label{eq:mettime}
e^{t}{}_{0} = f^{-1}\,, \qquad e^{t}{}_{i} = -f^2 W_i\,, \qquad e^{\alpha}{}_{0} = 0\,, \qquad e^{\alpha}{}_{i} = e^{-\l} \delta^{\alpha}_{i}\,,
\end{equation}
where we have  defined $f\equiv e^{\varphi}$ and $W_\a = e^{2\vf- \l} B_\a$. We write $\m = (t, \alpha)$ for the curved indices and $a = (0, i)$ for the flat ones, respectively. We
also require that all functions occurring in the metric (\ref{eq:mettime}) are independent of the coordinate $t$. Taking all these things into account, the components of the spin connection $\omega_{abc}$ in the flat basis read as follows,
\bea
\omega_{00i} &=& - e^{-\l}\,f^{-1}\partial_i f
\;,\nonumber\\[.1truecm]
\omega_{0ij} &=& -\omega_{ij0} ~=~ f\,e^{-2\l}\,\partial_{[i}\left(W_{j]}e^\sigma f^{-2}\right)
\;,\nonumber\\[.1truecm]
\omega_{ijk} &=& 2e^{-\l}\,\delta_{i[j}\partial_{k]}\l
\;.
\label{eq:omega}
\eea
Following \cite{Deger:2013yla}, it can be shown that the existence of a timelike Killing spinor leads to the
 following relations between the auxiliary fields $V_\mu$, $S$ and the metric functions
\bea
V_0&=& \frac12\, \epsilon^{ij}\,\omega_{ij0}\;,
\label{V00}\\[.5ex]
V_1 - i V_2 &=& \rmi e^{-\l}\,\partial_z\left(\varphi-\l+ic \right)
\;,\label{Vm}\\[.5ex]
S  &=&  \rmi  e^{-\l-ic}\,\partial_z\left(\varphi+\l-ic \right)
\;,
\label{eq:S}\\[.5ex]
\epsilon^{ij} \partial_i B_j &=& -2 V_0\, e^{2\l-\varphi}\;,
\label{eq:B}
\eea
where $c(x,y)$ is an arbitrary  time-independent real function and $z=x+iy$ denotes the complex coordinates.

At this stage we have paved the way for constructing supersymmetric background solutions by exploiting the Killing spinor identities. Making  an ansatz for the vector field $V_{\mu}$ we can now solve eqs.~(\ref{V00})--(\ref{eq:B}) and determine the metric functions $\l$ and $\varphi$. Following the same logic in \cite{Deger:2013yla}, we now look for solutions for with the following field configuration
\begin{equation}
S = \Lambda\,, \qquad \qquad V_a = \mbox{const}\,, \qquad \qquad V_2 = 0\,, \qquad \qquad c=0\,.
\label{config}
\end{equation}
With these choices, the non-vanishing components of the spin connection given in eq.~(\ref{eq:omega}) in a flat basis read as follows
\bea
\omega_{002} &=& -(\Lambda + V_1) \;, \qquad \omega_{112}~ = ~\Lambda - V_1\ ,
\nn\\[.1truecm]
\omega_{120}&=&\omega_{201}~=~-\omega_{012}~=~V_0\ .
\label{omegaconst}
\eea
Note that, by setting $V_{2}=c=0$,  we can solve for~$\l$ and $\vf$ using eqs.~(\ref{Vm}) and (\ref{eq:S}) and their integrability conditions. Furthermore, $B_{y}$ can be set to zero by a gauge choice. As a result, we obtain the following differential equations for the functions
$\varphi,$ $\l$ and $B_{x}$
\begin{eqnarray}
e^{-\l}\partial_{y}\varphi & = & V_{1}+\Lambda, \label{SusyConstraints1}\\[.1truecm]
e^{-\l}\partial_{y}\l & = & \Lambda-V_{1},\label{SusyConstraints2}\\[.1truecm]
\partial_y B_{x} & = & 2V_{0}\,e^{2\l-\varphi},
\label{SusyConstraints3}
\end{eqnarray}
with $\partial_{x}\varphi=\partial_{x}\l=0$.

It is worth emphasizing that so far we have not used the equations of motion, we have only considered the constraints that follow from supersymmetry. The solutions of eqs.~(\ref{SusyConstraints1})--(\ref{SusyConstraints3}) will bifurcate depending on the value of the vector component $V_{1}$. In the next subsection we will classify the supersymmetric solutions of the CNMG Lagrangian (\ref{NMG11}) with respect to the value of this vector field component by imposing the field equations.

\subsection{Classification of Supersymmetric Background Solutions}

In this subsection, we first integrate the differential equations (\ref{SusyConstraints1})--(\ref{SusyConstraints3}) depending on the different values of the vector field components $V_a$, which yields the metric functions $\l$ and $ \varphi$. Next, we impose the field equations and
 determine the couplings. The results for the different cases are given in three subsubsections.
  For the convenience of the reader, we have summarized all supersymmetric background solutions  allowed by the theory (\ref{NMG11}) in  Table \ref{table:1}.
\begin{table}[ht!]
\begin{center}
\begin{tabular}{ |c|c|c|c|c|c| }
\hline
& $V^2$ & $V_0$ & $V_1$ & Equation & Solution of STMG?  \\
\hline
Round $AdS_3$ & 0 & 0 & 0 & \ref{RoundAdSMetric} &  {\ding{51}} \\
\hline
$AdS_{2}\times \mathbb{R}$ &$ >0 $& $0$ & $\L$ & \ref{AdS2RMetric} & \xmark \\
\hline
Null-Warped $AdS_3$ & 0 & $\pm \L$ & $\L$ & \ref{nullwarped} & \cmark \\
\hline
Spacelike Squashed $AdS_3$ & $>0$ & $<\L$ & $\L$ & \ref{SpacelikeSquashed}& \cmark \\
\hline
Timelike Streched $AdS_3$ & $< 0$ & $> \L$ & $\L$ & \ref{tstreched}& \cmark \\
\hline
$AdS_3$ pp-wave & 0 & $V_0$ & $\varepsilon V_0$ & \ref{AdSppwave} & \cmark \\
\hline
Lifshitz & $>0$ & $0$ & $\neq 0$ and $\neq \L$ & \ref{LifMetric} & \xmark \\
\hline
\end{tabular}
\end{center}
\caption{Classification of supersymmetric background solutions of the $\cN = (1,1)$ CNMG. The solutions are classified with respect to the values of the components of the auxiliary vector $V_a$, and compared with the solutions of the $\cN = (1,1)$ TMG (STMG). }
\label{table:1}
\end{table}

\subsubsection{The case $V_1 = 0$}

 We start with the simplest case, i.e.~$V_1 = 0$. The supersymmetry constraint equations (\ref{SusyConstraints1})--(\ref{SusyConstraints3})  yield
\begin{equation}
\l=-\log(-\Lambda y),\qquad \quad \varphi=\log(-\frac{1}{\Lambda y}),\qquad \quad B_{x}=-\frac{2V_{0}}{\Lambda}\log(-\Lambda y).
\end{equation}
The vector equation (\ref{fieldequations}) then implies $V_{0}=0$ for $\Lambda\neq0$. Finally, from the scalar equation we fix $M$ to be
\begin{equation}
M=-\frac{\Lambda^{3}}{6m^{2}}+\Lambda\s .
\end{equation}
Thus, the metric becomes
\begin{equation}
ds^{2}=\frac{l^{2}}{y^{2}}\,(-dt^{2}+dx^{2}+dy^{2})\,,
\label{RoundAdSMetric}
\end{equation}
which describes the \textbf{round $\mathbf{AdS_3}$ spacetime} with $l=-\frac{1}{\L}$, see Table \ref{table:1}.
%After a coordinate transformation $y=1/r$
%\begin{equation}
%ds^{2}=l^{2}(-r^{2}dt^{2}+\dfrac{dr^{2}}{r^{2}}+dx^{2})
%\end{equation}

\subsubsection{The case $V_1 = \L\ne 0$}

For $V_{1}=\Lambda$, we obtain
\begin{equation}
\l=0,\qquad \qquad \varphi=2\Lambda y,\qquad \qquad B_{x}=-\frac{V_{0}}{\Lambda}e^{-2\Lambda y} \,.
\end{equation}
The vector and the scalar field equation lead to the following subclasses {\bf A, B}{ and {\bf C} which we describe below.

\subsubsection*{A. $V_{0}=0$, \hspace{0.2cm} $\L=-2 \sqrt{\dfrac{m^{2} \s}{7}}$, \hspace{0.2cm} $M=\dfrac{7 \L^{3}}{12 m^{2}}+\L \s$ }

With this choice of parameters the metric reads
\begin{equation}
ds^{2}=-e^{4\Lambda y}dt^{2}+dx^{2}+dy^{2}\,.
\end{equation}
After a simple coordinate transformation $y=\dfrac{\log{r}}{2\L}$, $x= \dfrac{x^{\prime}}{2 \L}$
the metric is brought into the following form
\begin{equation}
ds^{2}=\dfrac{l^{2}}{4}(- r^{2} dt^{2}+\dfrac{dr^{2}}{r^{2}}+dx^{2})\,.
\label{AdS2RMetric}
\end{equation}
which is $\mathbf{{AdS_{2} \times \mathbb{R}}}$. This background also appeared in the bosonic version of NMG, although
given in different coordinates \cite{Bergshoeff:2009aq, Clement:2009gq}.

\subsubsection*{B. $V_{0}=\pm\Lambda$, \hspace{0.2cm} $\L= -\sqrt{\dfrac{-2 m^{2}\s}{7}}$, \hspace{0.2cm} $M=-\dfrac{\L^{3}}{6 m^{2}}+\L \s$ }

This choice  of parameters leads to the metric
\bea
ds^2 = - e^{4 \L y} dt^2 \pm 2 e^{2 \L y} dt dx + dy^2 \,.\label{eq:nwarped}
\label{NullWarpedAdS}
\eea
Performing a coordinate transformation
\begin{equation}
y=l \log u,\qquad \qquad t=l x^{-},\qquad \qquad x=\pm \dfrac{l x^{+}}{2},
\end{equation}
the metric (\ref{eq:nwarped}) can be put into the  more familiar form \cite{Anninos:2008fx}
\begin{equation}
ds^{2}=l^{2}\left[\dfrac{du^{2}+dx^{+}dx^{-}}{u^{2}}-\left(\dfrac{dx^{-}}{u^{2}}\right)^{2}\right] ,\label{nullwarped}
\end{equation}
which is \textbf{null warped $\mathbf {AdS_3}$}.

\subsubsection*{C. $V_{0}=\pm \sqrt{\dfrac{7\Lambda^{2}-4 m^{2} \sigma}{21}}$, \hspace{0.2cm}  $M=-\dfrac{\L^{3}}{3 m^{2}}+ \dfrac{8\L \s}{7}$}

Using these values for the parameters and  fixing the value of $V_{0}$ we deduce from  the vector equation that
\bea
ds^2 = \frac{V^2}{\L^2} \Big( dx + \frac{V_0 \L}{V^2}\, e^{2 \L y}\, dt \Big)^2 - \frac{\L^2}{V^2} \,e^{4 \L y}\, dt^2 + dy^2\, .
\label{V2Form}
\eea
After making a coordinate transformation $\dfrac{V_{0}\L}{V^{2}}e^{2\L y}=\dfrac{1}{z}$, the metric reads
\begin{equation}
ds^2 = \frac{V^2}{\L^2} \Big( dx + \frac{dt}{z} \Big)^2 - \frac{1}{z^{2}}\, \frac{V^{2}}{\L^{2}}\, \dfrac{dt^2}{\nu^{2}} + \dfrac{dy^2}{4\L^{2}z^{2}}\,,
\end{equation}
where $\n^2 = 1 - \frac{V^2}{\L^2}  < 1$.

This is not yet the end of the story for this subclass: provided that $V^2 > 0$, which implies $7\L^{2}+2m^{2}\s > 0$, we have $1>\nu^{2}>0$. By making a coordinate transformation
\begin{equation}
x= \dfrac{x^{\prime}\nu}{2 V}\,, \qquad \qquad t=\dfrac{t^{\prime}\nu}{2 V}\,,
\end{equation}
the metric (\ref{V2Form}) can be cast into the following form
\bea
ds^2 &=& \frac{l^2}{4} \Big[ \frac{ - dt^2 + dz^2 }{z^2} + \n^2 \Big( dx + \frac{dt}{z} \Big)^2 \Big]\, ,
\label{SpacelikeSquashed}
\eea
which is the metric of \textbf{spacelike squashed $\mathbf{AdS_3}$} with  squashing parameter $\n^2$.

For $V^2 < 0$, i.e. $7\L^{2}+2m^{2}\s < 0$,  we perform  a coordinate transformation
\begin{equation}
x= \dfrac{x^{\prime}}{2}\sqrt{\dfrac{-\nu^{2}}{V^{2}}}\,, \qquad \qquad t=\dfrac{t^{\prime}}{2}\sqrt{\dfrac{-\nu^{2}}{V^{2}}}\,,
\end{equation}
after which the metric  (\ref{V2Form}) can be written in the following form
\bea
ds^2 = \frac{l^2}{4} \Big[ \frac{ dt^2 + dz^2 }{z^2} - \n^2 \Big( dx + \frac{dt}{z} \Big)^2 \Big]\label{tstreched}\,,
\eea
where $\n^2 > 1$. The metric (\ref{tstreched}) is one of the incarnations of the \textbf{timelike stretched~$\mathbf{AdS_3}$}
background.

\subsubsection{The case $V_{1}\neq\Lambda$ and $V_{1}\neq0$}

This class of solutions have $V_{1}\neq\Lambda$ and $V_{1}\neq0$. The calculation of the metric functions follows the computations performed in the previous subsubsections with the extra definitions
\begin{equation}
\sigma=-\log(z),\qquad \quad \varphi=\log(z^{\alpha}),\qquad \quad B_{x}=-\frac{V_{0}}{V_{1}}z^{-(1+\alpha)},
\label{V1neqL}
\end{equation}
where
\begin{equation}
z \equiv (V_{1}-\Lambda)y,\qquad \qquad \alpha\equiv\frac{V_{1}+\Lambda}{V_{1}-\Lambda}\label{alpha} \,.
\end{equation}
Using the components of the vector equation, we find
\begin{equation}
V_{0}(V_{0}^{2}-V_{1}^{2})(V_{1}-\Lambda)=0 \label{V1neqLeq} \,.
\end{equation}
From eq.~(\ref{V1neqLeq}) it is straightforward to see that this subclass has two different branches, i.e. $V_{0}=0$ and $V_{1}=\varepsilon V_{0}$ with $\varepsilon^{2}=1$. We will discuss these two branches as separate cases
{\bf A} and {\bf B} below,

\subsubsection*{A. $V_{1}=\varepsilon V_{0}\,, \varepsilon=\pm 1$\,, \hspace{0.2cm}
$V_{0}=-\varepsilon\L\pm\sqrt{\dfrac{\L^{2}-2m^{2}\s}{2}}$ }

With this choice of parameters  the vector equation gives rise to
\begin{equation}
2V_{0}^{2}+4\varepsilon V_{0}+\Lambda^{2}+2m^{2}\s=0 \,.
\end{equation}
The parameter $M$ can be solved by using the field equation for $A$ as follows,
\begin{equation}
M=\frac{-\Lambda^{3}}{6m^{2}} + \Lambda\s \,.
\end{equation}
Plugging in the metric functions, we obtain the following expression for the metric
\[
ds^{2}=-z^{2\alpha}(-dt+2\varepsilon z^{-1-\alpha}dx)dt+\frac{1}{(V_{1}-\Lambda)^{2}}\frac{dz^{2}}{z^{2}}\,.
\]
Performing the  coordinate transformation \cite{Deger:2013yla}
\bea
z = u^{\frac{(\L - V_1)}{\L}}\,, \qquad \qquad t = l x^- \,, \qquad \qquad x =\frac{ \varepsilon l x^+}{2} \,,
\eea
this metric can be written as follows
\begin{equation}
ds^{2}= l^{2}\left[\dfrac{du^{2}+dx^{+}dx^{-}}{u^{2}}-u^{2(\frac{\L-V_{1}}{\L})}\left(\frac{dx^{-}}{u^{2}}\right)^{2}\right]\,.
\label{AdSppwave}
\end{equation}
This is the metric of a  $\mathbf{AdS_{3}}$ \textbf{pp-wave}. Note that the limit $V_{1}\rightarrow\L$ is  well defined and
gives rise to the \emph{minus} null warped $AdS_{3}$ metric of eq.~(\ref{nullwarped}), as expected.

\subsubsection*{B. $V_0 = 0$, \hspace{0.2cm} $V_{1}=\dfrac{\a+1}{\a-1}$, \hspace{0.2cm} $M=\frac{\Lambda(9V_{1}^{2}-2\Lambda^{2})}{12m^{2}}+\Lambda\s$}

The final spacetime we consider appears for $V_{0}=0$.  Rather than solving the vector equation for $V_{1}$ as we did in the previous cases, we set  $V_{1}=\dfrac{\a+1}{\a-1}$ using eq.~(\ref{alpha}). The field equations further imply that
\begin{equation}
(1 - 14 \a - 7 \a^2) \L^2 +
4 m^2 (-1 + \a)^2 \s=0,
\end{equation}
whose solution is given by
\begin{equation}
\L=-\sqrt{\dfrac{4m^{2}\s(\a-1)^{2}}{ (1 - 14 \a - 7 \a^2)}}\,.
 \end{equation}
 Here, we would like to restrict our attention to $\a < 0$, as $\a$ will be the minus of the Lifshitz exponent, thus giving rise to spacetimes with positive Lifshitz exponent
\begin{description}
\item{(1)}\ \  $\a< \dfrac{1}{7} (-7 - 2 \sqrt{14})$ then $m^{2}\s>0$,

\item{(2)}\ \ $\dfrac{1}{7} (-7 - 2 \sqrt{14})< \a <0 $ then $m^{2}\s<0$,

\end{description}
Provided that the vector field components are chosen as discussed, we obtain the \textbf{Lifshitz} metric
\begin{equation}
ds^2 = l_L^2 \Big[ - y^{2\a} dt^{2} + \frac{1}{y^2} (dx^2 + dy^2 ) \Big] \,,
\end{equation}
where $l_L$ is the Lifshitz radius which is defined as
\bea
l_L^2 = \frac{1}{ (V_1 - \L )^2} \,.
\eea
We have redefined $t$ as $t \rightarrow (V_1 - \L)^{2\a + 2} t$. Note that in the limit $V_{1}\rightarrow 0$ one obtains the round $AdS_{3}$ metric given in eq.~\eqref{RoundAdSMetric}. Taking $y = \frac1{r}$ gives the metric in the standard form
\bea
ds^2 = l_L^2 \Big( - r^{-2\a} dt^2 + r^2 dx^2 + \frac1{r^2} dr^2 \Big) \,,
\label{LifMetric}
\eea
where $l_L^2$ and $V_1$ are given in terms of $\a$ and $\L$ as\,\footnote{Note that the standard Lifshitz exponent $z$ in the literature is given by $z=-\a$.}
\bea
l_L^2 = \Big( \frac{\a - 1}{2\L} \Big)^2.
\eea

As shown in \cite{Deger:2013yla}, all the supersymmetric backgrounds that we have found in this section except the $AdS_3$ metric preserve $1/4$ of the supersymmetries.

\section{Supersymmetric Black Holes}{\label{BlackHoles}}

In this section, we discuss the supersymmetry aspects of black hole solutions of CNMG  in a $AdS_3$ or Lifshitz background. The existence of a Killing spinor is highly restricted due to the global requirement that the angular coordinate $\phi$ should be periodic. As shown in \cite{Deger:2013yla}, the spacelike squashed $AdS_3$ solution can be interpreted as an extremal black hole upon making a coordinate transformation. Therefore, in this section we will discuss three specific cases of black hole solutions. We start our discussion in subsection \ref{ss: HairyBTZ} with a generalization of the BTZ black hole, and show that the periodicity condition implies the extremality of the black hole. In the next subsection we  investigate the `logarithmic' black hole given in \cite{Clement:2009ka}, and show that, the logarithmic black hole is also supersymmetric. Finally, in a third subsection  we investigate the possible black holes in a  Lifshitz background.

\subsection{The Rotating Hairy BTZ Black Hole and its Killing Spinors}{\label{ss: HairyBTZ}}

The CNMG Lagrangian (\ref{NMG11}) admits the following rotating black hole solution \cite{Giribet:2009qz}

\begin{equation}
ds^{2}=-N^2F^2dt^{2}+\frac{dr^{2}}{F^2}+r^{2}\left(  d\phi+N^{\phi}dt\right)
^{2}\ , \label{Rotating black hole metric}%
\end{equation}
where $N$, $N^{\phi}$ and $F$ are functions of the radial coordinate $r$,
given by
\begin{align}
N^2  &  =\left[  1 + \frac{b}{4H}\left(  1-\Xi^{\frac{1}{2}}\right)  \right]
^{2}\ ,\nonumber\\
N^{\phi}  &  =-\frac{\cJ}{2\cM r^{2}}\left( \cM - b H\right)  \ ,\label{Ns&F}\\
F^2  &  =\frac{H^{2}}{r^{2}}\left[  H^2 +\frac{b}{2}\left(
1+\Xi^{\frac{1}{2}}\right)  H+\frac{b^{2}}{16}\left(  1-\Xi^{\frac{1}{2}%
}\right)  ^{2}- \cM\ \Xi^{\frac{1}{2}}\right]  \ ,\nonumber
\end{align}
and
\begin{equation}
H=\left[  r^{2}- \frac12 \cM \left(  1-\Xi^{\frac{1}{2}}\right)  -\frac{b^{2}
}{16}\left(  1-\Xi^{\frac{1}{2}}\right)  ^{2}\right]  ^{\frac{1}{2}}\ .
\label{H}%
\end{equation}
where we have set the $AdS_3$ radius $l = 1$. Here $\Xi:=1-\cJ^{2}/\cM^{2}$, and the rotation parameter~$\cJ/\cM$ is bounded in terms
of the AdS radius according to
\begin{equation}
-1\leq \cJ/\cM \leq 1\,.
 \end{equation}
 The parameter $b$ is the gravitational hair, and for $b=0$ one recovers the BTZ black hole \cite{Banados:1992wn}.  Since we impose the global requirement that $\phi$ should be periodic, i.e.~$0 \leq \phi \leq 2\pi$, the vacuum of the BTZ black hole with gravitational hair, defined by $\cM = \cJ = b = 0$, admits only two Killing spinors. In order to see that, we consider the Killing spinor equations (\ref{KSAds}). Since the equations for $\x_1$ and $\zeta_2$ enforce exponential solutions for $\x_1$ and $\zeta_2$, we cannot find a solution for $\x_2$ and $\zeta_1$ that is periodic in $\phi$. Therefore, finding a  periodic solution requires setting $\x_1 = \zeta_2 = 0$. This  implies that only two of the solutions of equations (\ref{KSAds}) are valid.

 Introducing the following orthonormal frame for the metric
\bea
e^0 = NF dt \,, \qquad \qquad e^1 =  rd\phi + r N^{\phi} dt  \,, \qquad \qquad e^2 = F^{-1} dr \,,
\eea
the spin-connection components are given by
\bea
\o_{01} &=& \frac{1}{2}\frac{r N^{\phi\prime}}{F N} dr \,,\qquad \o_{02} =\left( -F N F^{\prime}+\frac{ r^2  N^{\phi}  N^{\phi\prime}}{2 N}
-F^{2} N^{\prime}\right) dt + \frac{r^2 N^{\phi\prime}}{2 N} d\phi \,,\nn\\[.1truecm]
 \o_{12} &=&  \frac{1}{2} F (2 N^{\phi} + r N^{\phi\prime})dt+ F d\phi \,,
\eea
and hence the Killing spinor equation reads
\bea
0 = d\e  &+& \frac12 \Big( - \frac{r N^{\phi \prime}}{2FN} \s_3 \e + \frac{1}{F} \s_3 \e^\star    \Big) dr +  \frac12 \Big( \frac{r^2 N^{\phi \prime}}{2N} \s_1 \e -  \rmi F \s_2 \e +  r \s_1 \e^\star   \Big) d\phi  \nn\\[.1truecm]
&+&\frac12 \Bigg[ \Big(- F N F^\prime + \frac{r^2 N^\phi N^{\phi \prime}}{2N} - F^2 N^\prime \Big) \s_1 \e  -  \rmi \Big(F N^\phi  + \frac12 r F N^\prime \Big) \s_2 \e \nn\\[.1truecm]
&& \qquad  +  \rmi NF \s_2 \e^\star  + r N^\phi \s_1 \e^\star  \Bigg] dt \,.
\eea
Decomposing the Dirac spinor into two Majorana spinors like in eq.~\eqref{DiracMajorana}, we obtain the following equations
\bea
0 &=& d \xi_1 + \frac{1}{4 N} \Big( N^{\phi}\, [2N(r - F) + r^2 N^{\phi\prime} ]  - F N ( - 2 N + 2 r F^\prime + r N^{\phi\prime} + 2 F N^\prime ) \Big) \xi_2\, dt \nn\\[.1truecm]
&& \qquad  + \frac{1}{4 N} \Big(  2 N (r - F)  + r^2   N^{\phi \prime}    \Big) \xi_2 \, d\phi +  \frac{1}{4 F N} \Big(2 N  - r  N^{\phi \prime}            \Big) \xi_1 \, dr \,,\nn\\[.1truecm]
0 &=& d \xi_2  + \frac{1}{4 N} \Big( N^\phi \, [2 N (r + F)  + r^2  N^{\phi \prime} ] + F N (-2 N  - 2 r F' + r N^{\phi \prime} - 2 F N^\prime )\Big) \xi_1 \, dt
\nn\\[.1truecm]
&& \qquad  + \frac1{4N} \Big( 2 N (r + F ) + r^2 N^{\phi \prime}    \Big) \xi_1 \, d\phi + \frac{1}{4 F N} \Big(-2 N  + r  N^{\phi \prime}   \Big) \xi_2 \, dr \,,\nn\\[.1truecm]
0 &=& d \zeta_1 + \frac{1}{4 N} \Big( N^\phi \, [-2 N (r + F )  + r^2  N^{\phi \prime} ] - F N (2 N  +2r F^\prime + r N^{\phi \prime} + 2 F N^\prime ) \Big) \zeta_2 \, dt \nn\\[.1truecm]
&& \qquad +  \frac1{4N} \Big( -2 N (r + F ) + r^2 N^{\phi \prime}    \Big) \zeta_2 \, d\phi   - \frac{1}{4 F N} \Big(2 N  + r  N^{\phi \prime}   \Big) \zeta_1 \, dr  \,,\nn\\[.1truecm]
0 &=& d \zeta_2 + \frac{1}{4 N} \Big( N^\phi \, [-2 N (r - F)  + r^2  N^{\phi \prime} ] + F N (2 N  - 2 r F' + r N^{\phi \prime} - 2 F N^\prime ) \Big) \zeta_1 \, dt \nn\\[.1truecm]
&& \qquad +  \frac1{4N} \Big( -2 N (r - F ) + r^2 N^{\phi \prime}    \Big) \zeta_1 \, d\phi   + \frac{1}{4 F N} \Big(2 N  + r  N^{\phi \prime}   \Big) \zeta_2 \, dr  \,.
\label{BHTKS}
\eea
From these equations it follows that for the  generic case not all  the $d\phi$ components can be set to zero, which is the requirement for finding a periodic Killing spinor. Therefore, we turn our attention to the extremal solutions with $\cM = |\cJ|$. For this case we find  the following Killing spinors that are periodic in $\phi$
\begin{description}
\item{(1)}\ \ \ {\underline{$\cM = - \cJ$}
	\begin{eqnarray}
	\xi_1 = \zeta_1 = \zeta_2 = 0 \,, \quad \qquad \x_2 = \frac{b+ \sqrt{-b^2 + 8J + 16 r^2}}{\sqrt{r}} \,,
	\end{eqnarray}
}
\item{(2)}\ \ \ {\underline{$\cM = \cJ$}
	\begin{eqnarray}
	\xi_1 = \xi_2 = \zeta_2 = 0 \,, \quad \qquad \zeta_1 = \frac{b+ \sqrt{-b^2 - 8J + 16 r^2}}{\sqrt{r}}  \,.
	\end{eqnarray}
}
\end{description}}
\noindent Note that for zero hair, i.e.~$b \rightarrow 0$, one re-obtains the Killing spinors for a BTZ black hole.

\subsection{The `Logarithmic' Black Hole}

The supersymmetric CNMG Lagrangian \eqref{NMG11} also admits the following so-called `logarithmic' black hole solution \cite{Clement:2009ka}
\bea
ds^2 = - \frac{4 \r^2}{l^2 f^2(\r)}\, dt^2 + f^2(\r) \Big( d\phi - \ve\, \frac{q\, l \ln[\frac{\r}{\r_0}]  }{f^2 (\r)}\,  dt      \Big)^2 + \frac{l^2}{4\r^2}\, d\r^2 \,,
\eea
where $q \leq 0$ and $0 < \phi < 2\pi$. The function $f^2 (\r)$ is defined by
\bea
f^2 (\r) = 2 \r + q\, l^2 \ln[\frac{\r}{\r_0}] \,,
\eea
and the parameter $\ve = \pm 1$ determines the direction of the rotation since
\bea
M = 2q \,, \qquad \qquad J = 2\,\ve\, l q \,.
\eea
Setting $q = 0$ and making the  coordinate transformation $\r = r^2 / 2$ we obtain  a  $AdS_3$ background with $\phi$ being periodic. This implies that  the background of the `logarithmic' black hole preserves only half of the supersymmetries like in the case of the  rotating hairy BTZ black hole in the previous subsection.

We now determine the explicit expressions for the Killing spinors. Introducing the following orthonormal frame for the metric
\bea
e^0 = \frac{2\r}{l f(\r)} \, dt \,, \qquad e^1 = f(\r) \, d\phi - \frac{l q \,\ve \ln[\frac{\r}{\r_0}]}{f(\r)} \, dt \,, \qquad e^2 = \frac{l}{2\r} d\r \,,
\eea
we find the following expressions for the spin-connection components
\bea
\o_{01} &=& - \frac{l^2 q\, \ve} {4 \r^2 f(\r)} \, \Big[ f(\r) - 2\r\, f'(\r) \ln[\frac{\r}{\r_0}]   \Big] \, d\r \,, \qquad \o_{12} = - \frac{q\, \ve}{f(\r)}\, dt + \frac{2 \r\, f'(\r)}{l}\, d\phi \,,\nn\\
\o_{02} &=& - \frac{1}{2 l^2\, \r\, f^2 (\r)} \Bigg( f(\r) \Big[8\r^2 - l^4 q^2 \ln[\frac{\r}{\r_0}]  \Big] + 2\r f'(\r) \Big[ -4\r^2 +l^4 q^2  \ln[\frac{\r}{\r_0}]   \Big] \Bigg) dt \nn\\
&&  - l q\, \e \, \Big( \frac{f(\r)}{2\r}  + \ln[\frac{\r}{\r_0}] \,f'(\r)    \Big)\, d\phi \,.
\eea
Using these expressions in the Killing spinor equation (\ref{KillingSpinor}), we find that the Killing spinors of the logarithmic black hole are given by
\begin{description}
\item{i.}\ \ \ {\underline{$\ve = 1$}
	\begin{eqnarray}
	\xi_1 = \xi_2 = \zeta_2 = 0 \,, \quad \qquad \zeta_1 = \sqrt{\frac{\r}{\r_0}} \Big(\frac{1}{2r + l^2 q \ln[\frac{\r}{\r_0}]} \Big)^{1/4} \,,
	\end{eqnarray}
}
\item{ii.}\ \ \ {\underline{$\ve = - 1$}
	\begin{eqnarray}
	\xi_1 = \zeta_1 = \zeta_2 = 0 \,, \quad \qquad \xi_2 = \sqrt{\frac{\r}{\r_0}} \Big(\frac{1}{2r + l^2 q \ln[\frac{\r}{\r_0}]} \Big)^{1/4}  \,.
	\end{eqnarray}
}
\end{description}

This result may be somewhat surprising considering the expectation that the only existing supersymmetric black hole in an $AdS_3$ background is an extremal BTZ black hole \cite{Bergshoeff:2010mf}. However, unlinke the rotating BTZ black hole, the "logarithmic" black hole does not have a non-extremal limit $J \neq M$. Thus, one cannot recover a static, non-supersymmetric black hole from the $J \rightarrow 0$ limit of the  "logarithmic" black hole. Therefore, this particular case evades the argument presented in \cite{Bergshoeff:2010mf}.\footnote{We thank Paul Townsend for a clarifying discussion on this exceptional case.}

\subsection{Searching For a Supersymmetric Lifshitz Black Hole}{\label{LBH}}

In this section, we briefly present our attempts to find a supersymmetric Lifshitz black hole. Following \cite{Lu:2012am}, we first try to saturate the BPS bound using the vector field $V_\m$, since it can, in principle, contribute as a massive vector hair. In order to do so, we consider the following metric ansatz
\bea
ds^2 = l_L^2 \Big(- a dt^2 + r^2 dx^2 + \frac1{f} dr^2           \Big) \,,
\label{LBHAnsatz}
\eea
where the functions $a$ and $f$ depend on the coordinate $r$ only. With this ansatz for the metric, one can show that the Killing spinor equation imposes the following constraint of these functions
\bea
\frac{a^\prime \sqrt{f}}{a} + \frac{2\sqrt{f}}{r} + 2 (\a -1)  = 0 \,.
\label{LBHConstraint}
\eea
Having obtained this constraint, we next turn to the vector equation~(\ref{fieldequations}). Using the metric ansatz (\ref{LBHAnsatz}), the $V_0$ and $V_2$ components of the vector equation are automatically satisfied, while the $V_1$ component reads
\bea
0 &=& (1+\a) \Big[ r^2 f a' + 2 a^2 \Big(- 8 f + r [2r ( -1 + 5 \a + \a^2 ) + 5 f' ] \Big) \nn\\[.1truecm]
&& \qquad \quad - r a \Big( ra'f' + 2f(-5a' + ra''  )            \Big)             \Big].
\label{veceqn1}
\eea
Imposing the Killing spinor constraint (\ref{LBHConstraint}) to simplify the vector equation, we obtain
\bea
-7r \sqrt{f} (\a - 1) - 11 f + r\Big( r ( 7\a - 2) + 3 f'   \Big)  = 0 \,.
\eea
As we wish to find a solution for $f$ which has a double root at $r = r_0$, which is a necessary condition for an extremal black hole, we need to be able to eliminate the $f$ terms in the vector equation. Using the fact that the Killing spinor constraint (\ref{LBHConstraint}) can be cast into
the following form
\bea
\sqrt{f} (1-\a) = -\frac12 \Big( \frac{a^\prime}{a}  + \frac2r  \Big) f \,,
\eea
 the vector equation can be written as
\bea
\frac72 r \Big( \frac{a^\prime}{a}  - \frac8{7r}  \Big) f + r\Big( r ( 7\a - 2) + 3 f'   \Big)  = 0 \,,
\eea
which has the following solution
\bea
a = r^{8/7} \,, \qquad f =  r^2 - r_0^2 \,.
\eea
However, using this equation in the Killing spinor constraint (\ref{LBHConstraint}), we find that $r_0 = 0$. A further check with the metric equation also imposes $r_0 = 0$. Therefore, although the Killing spinor equation allows the existence of a supersymmetric black hole, we find that the vector and metric equations are incompatible with that possibility.

Alternatively, one may try to start with a rotating Lifshitz black hole using the following metric ansatz
\bea
ds^2 = l_L^2 \left[- r^{-2\alpha} F(r) dt^2 + \Big( r dx + r^{-\alpha} G(r) dt \Big)^2 + \frac1{r^2 F(r)} dr^2          \right] \,,
\label{LBHAnsatzRot}
\eea
where $F(r)$ and $G(r)$ are arbitrary functions that depend on the coordinate $r$ only. In this case the Killing spinor equation  constrains the function $F(r)$ to be of the form
\bea
F(r) = 1 + a r^{-2 + 2\a} \,,
\label{Br}
\eea
where $a$ is a constant. Furthermore, the vector equation constraints the function $G(r)$ via the following differential equation
\bea
21 r^4 G^{\prime 2} - 42 r^3 (\a + 1)  G G^\prime + 21 r^2 (1 + \a)^2 G^2 + 4 a r^{2\a} (6\a - 11) = 0\,.
\eea
Using the solutions of this differential equation, along with the expression (\ref{Br}) in the gravity equation, we find that it takes us back to the Lifshitz background, not allowing a rotating black hole solution.

The result of this subsection is somewhat expected, considering the fact that for the only rotating Lifshitz solution known to us \cite{Sarioglu:2011vz}, the couplings are determined by using a stationary Lifshitz spacetime which has a rotation term. This is not allowed by the given matter configuration of the $\cN = (1,1)$ CNMG theory.

Finally, we would like to comment that as our attempts to find a supersymmetric Lifshitz black hole has failed with the parity-even theory under our consideration (\ref{NMG11}), one may consider to modify the CNMG by adding a parity violating Lorentz-Chern-Simons term, which gives rise to the so-called $\cN = (1,1)$ Generaized Massive Gravity (SGMG) \cite{Alkac:2014hwa}. In that case, we found that the vector equation is modified in such a way that the Lifshitz background is no longer a solution with the field configuration given in (\ref{config}).

\section{Conclusions}{\label{Conclusions}}

Using the off-shell Killing spinor analysis, we have investigated in this work the supersymmetric backgrounds of the $\cN = (1,1)$ CNMG model given by the Lagrangian \eqref{NMG11}. The background solutions are classified according to the norm of the Killing vector constructed out of Killing spinors. There are two cases only. First of all, when the Killing vector is null, see section 3, the $\cN = (1,1)$ analysis reduces to that of the $\cN = 1$ CNMG model, since the null choice forces the
auxiliary massive vector $V_\m$ and the auxiliary pseudo-scalar~$B$ to vanish. Therefore, the solution is of the pp-wave type which preserves  $1/4$ of the supersymmetries. In the $AdS_3$ limit, there is a supersymmetry enhancement, and the $AdS_3$ solution is maximally supersymmetric.

As a second case, in section 4 we investigated the case that the Killing vector is taken to be timelike.
 In particular, we did consider a special class of solutions in which the pseudo-scalar $B$ vanishes. In that case all the supersymmetric solutions can be classified in terms of the components $V_a$ of the massive vector in the flat basis. A subclass of these solutions, with different parameters, are also  solutions of the supersymmetric TMG model,
 see Table \ref{table:1}.  In addition to these solutions, we found that the $\cN =(1,1)$ CNMG model possesses Lifshitz and $AdS_2 \times {\mathbb{R}}$ solutions. All these background solutions  preserve $1/4$ of the superymmetries.

Next, in section 5 we  investigated three cases of  black hole solutions in a $AdS_3$ or Lifshitz background. In the case of $AdS_3$, we studied the rotating hairy BTZ black hole in subsection 5.1  and the logarithmic black hole in
subsection 5.2. We found that in general the rotating hairy BTZ black hole  is not supersymmetric due to the fact that the periodicity condition on the $\phi$ coordinate and the periodic Killing spinors only arise when the black hole is extremal. In the  case of the  logarithmic black hole, we found that only the extremal  black hole solution exists, which is supersymmetric by its own nature. Finally, we  analyzed the conditions for the existence of a supersymmetric Lifshitz black hole, and showed that it does not exist given the field configuration of the $\cN = (1,1)$ CNMG model.

There are numerous directions one can consider for future study. An intriguing problem is to find a supersymmetric Lifshitz black hole. Although our trials with the current model has failed, it is natural to consider different approaches. For instance, one could  saturate the BPS bound with a $U(1)$ charge. This can be achieved by coupling the $\cN = (1,1)$ CNMG model to an off-shell vector multiplet and repeat the analysis presented in this paper.

Finally, we would like to mention that the same procedure that we presented in this paper can be applied to the $\cN = (2,0)$ CNMG model. This model has a different field content consisting  of two auxiliary vectors and a real scalar as well as the graviton and the gravitino. Given that the $\cN=(2,0)$ theory with matter couplings has new supersymmetric solutions \cite{Howe:1995zm}, we would expect that the $\cN = (2,0)$ CNMG model exhibits different supersymmetric solutions. Therefore, it would be interesting to see what the consequences of the different field content is for the supersymmetric solutions of the model.

\section*{Acknowledgements}

We thank Ozgur Sarioglu and Paul Townsend for useful comments and discussions. The research of L.B. and M.O are supported by the Dutch stichting voor Fundamenteel Onderzoek der Materie (FOM). G.A. acknowledges support by a grant of the Dutch Academy of Sciences (KNAW).  D.O.D acknowledges financial support and kind hospitality at the Van Swinderen Institute while part of this paper
was conceived. D.O.D is also partially supported by the Scientific and Technological Research Council of Turkey (T{\"U}B\.{I}TAK) Grant No.113F034.


\begin{thebibliography}{99}
\bibliographystyle{utphys}

\bibitem{Gates:1983nr}
S.~J.~Gates, M.~T.~Grisaru, M.~Rocek and W.~Siegel,
``Superspace Or One Thousand and One Lessons in Supersymmetry,''
Front.\ Phys.\  {\bf 58}, 1 (1983)
[hep-th/0108200].

\bibitem{Siegel:1979fr}
W.~Siegel,
``Unextended Superfields in Extended Supersymmetry,''
Nucl.\ Phys.\ B {\bf 156}, 135 (1979).

\bibitem{Brown:1979ma}
M.~Brown and S.~J.~Gates, Jr.,
``Superspace Bianchi Identities and the Supercovariant Derivative,''
Annals Phys.\  {\bf 122}, 443 (1979).
%%CITATION = APNYA,122,443;%%

%\cite{Uematsu:1984zy}
\bibitem{Uematsu:1984zy}
T.~Uematsu,
``Structure of $N=1$ Conformal and Poincare Supergravity in (1+1)-dimensions and (2+1)-dimensions,''
Z.\ Phys.\ C {\bf 29} (1985) 143;

\bibitem{Uematsu:1986de}
T.~Uematsu,
``Constraints and Actions in Two-dimensional and Three-dimensional $N=1$ Conformal Supergravity,''
Z.\ Phys.\ C {\bf 32}, 33 (1986).

  %\cite{Deser:1981wh}
  \bibitem{Deser:1981wh}
  S.~Deser, R.~Jackiw and S.~Templeton,
  ``Topologically massive gauge theories'',
  Annals Phys.\  {\bf 140} (1982) 372
  [Erratum-ibid.\  {\bf 185} (1988\ APNYA,281,409-449.2000) 406.1988\ APNYA,281,409].
  %%CITATION = APNYA,281,409;%%


\bibitem{DeserKay}
S. Deser and J.H. Kay,
`` Topologically massive supergravity",
Phys. Lett. {\bf B120} (1983) 97.
%%CITATION = PHLTA,B120,97;%%

\bibitem{Deser}
S. Deser,
`` Cosmological topological supergravity ",
in ``Quantum Theory of Gravity,'' ed. S.M. Christensen (Adam Hilger, London,
1984).
%%CITATION = PRINT-82-0692-BRANDEIS-;%%

\bibitem{Gibbons:2008vi}
G.~W.~Gibbons, C.~N.~Pope and E.~Sezgin,
``The General Supersymmetric Solution of Topologically Massive Supergravity,''
Class.\ Quant.\ Grav.\  {\bf 25}, 205005 (2008)
[arXiv:0807.2613 [hep-th]].

%\cite{Bergshoeff:2009hq}
\bibitem{Bergshoeff:2009hq}
E.~A.~Bergshoeff, O.~Hohm and P.~K.~Townsend,
``Massive Gravity in Three Dimensions,''
Phys.\ Rev.\ Lett.\  {\bf 102} (2009) 201301
[arXiv:0901.1766 [hep-th]].
%%CITATION = ARXIV:0901.1766;%%
%386 citations counted in INSPIRE as of 29 juin 2015


\bibitem{Andringa:2009yc}
R.~Andringa, E.~A.~Bergshoeff, M.~de Roo, O.~Hohm, E.~Sezgin and P.~K.~Townsend,
``Massive 3D Supergravity,''
Class.\ Quant.\ Grav.\  {\bf 27}, 025010 (2010)
[arXiv:0907.4658 [hep-th]].

\bibitem{Bergshoeff:2010mf}
E.~A.~Bergshoeff, O.~Hohm, J.~Rosseel, E.~Sezgin and P.~K.~Townsend,
``More on Massive 3D Supergravity,''
Class.\ Quant.\ Grav.\  {\bf 28}, 015002 (2011)
[arXiv:1005.3952 [hep-th]].


\bibitem{Alkac:2014hwa}
G.~Alkac, L.~Basanisi, E.~A.~Bergshoeff, M.~Ozkan and E.~Sezgin,
``Massive N=2 Supergravity in Three Dimensions,''
arXiv:1412.3118 [hep-th].

%\cite{Kuzenko:2015jda}
\bibitem{Kuzenko:2015jda}
  S.~M.~Kuzenko, J.~Novak and G.~Tartaglino-Mazzucchelli,
  ``Higher derivative couplings and massive supergravity in three dimensions,''  arXiv:1506.09063 [hep-th].  %%CITATION = ARXIV:1506.09063;%%




\bibitem{Deger:2013yla}
N.~S.~Deger, A.~Kaya, H.~Samtleben and E.~Sezgin,
``Supersymmetric Warped AdS in Extended Topologically Massive Supergravity,''
Nucl.\ Phys.\ B {\bf 884}, 106 (2014)
[arXiv:1311.4583 [hep-th]].


\bibitem{Kuzenko:2013uya} 
S.~M.~Kuzenko, U.~Lindstrom, M.~Rocek, I.~Sachs and G.~Tartaglino-Mazzucchelli,
``Three-dimensional $\mathcal{N} =$ 2 supergravity theories: From superspace to components,''
Phys.\ Rev.\ D {\bf 89}, no. 8, 085028 (2014)
[arXiv:1312.4267 [hep-th]].

\bibitem{Giribet:2009qz}
G.~Giribet, J.~Oliva, D.~Tempo and R.~Troncoso,
``Microscopic entropy of the three-dimensional rotating black hole of BHT massive gravity,''
Phys.\ Rev.\ D {\bf 80} (2009) 124046
[arXiv:0909.2564 [hep-th]].


\bibitem{Banados:1992wn}
M.~Banados, C.~Teitelboim and J.~Zanelli,
``The Black hole in three-dimensional space-time,''
Phys.\ Rev.\ Lett.\  {\bf 69}, 1849 (1992)
[hep-th/9204099].

\bibitem{Clement:2009ka}
G.~Clement,
``Black holes with a null Killing vector in new massive gravity in three dimensions,''
Class.\ Quant.\ Grav.\  {\bf 26}, 165002 (2009)
[arXiv:0905.0553 [hep-th]].

\bibitem{Lu:2012am}
H.~Lu and Z.~L.~Wang,
``Supersymmetric Asymptotic AdS and Lifshitz Solutions in Einstein-Weyl and Conformal Supergravities,''
JHEP {\bf 1208}, 012 (2012)
[arXiv:1205.2092 [hep-th]].


%\bibitem{VanProeyen:1999ni}
%A.~Van Proeyen,
%``Tools for supersymmetry,''
%hep-th/9910030.

%\cite{Anninos:2008fx}
\bibitem{Anninos:2008fx}
D.~Anninos, W.~Li, M.~Padi, W.~Song and A.~Strominger,
``Warped AdS(3) Black Holes,''
JHEP {\bf 0903}, 130 (2009)
[arXiv:0807.3040 [hep-th]].
%%CITATION = ARXIV:0807.3040;%%
%189 citations counted in INSPIRE as of 26 juin 2015




\bibitem{AyonBeato:2009yq}
E.~Ayon-Beato, G.~Giribet and M.~Hassaine,
``Bending AdS Waves with New Massive Gravity,''
JHEP {\bf 0905}, 029 (2009)
[arXiv:0904.0668 [hep-th]].

\bibitem{Bergshoeff:2009aq}
E.~A.~Bergshoeff, O.~Hohm and P.~K.~Townsend,
``More on Massive 3D Gravity,''
Phys.\ Rev.\ D {\bf 79}, 124042 (2009)
[arXiv:0905.1259 [hep-th]].
%%CITATION = ARXIV:0905.1259;%%
%196 citations counted in INSPIRE as of 26 Jun 2015

\bibitem{Clement:2009gq}
G.~Clement,
``Warped AdS(3) black holes in new massive gravity,''
Class.\ Quant.\ Grav.\  {\bf 26}, 105015 (2009)
[arXiv:0902.4634 [hep-th]].


\bibitem{Sarioglu:2011vz}
O.~Sarioglu,
``Stationary Lifshitz black holes of $R^2$-corrected gravity theory,''
Phys.\ Rev.\ D {\bf 84}, 127501 (2011)
[arXiv:1109.4721 [hep-th]].

\bibitem{Howe:1995zm} 
P.~S.~Howe, J.~M.~Izquierdo, G.~Papadopoulos and P.~K.~Townsend,
``New supergravities with central charges and Killing spinors in (2+1)-dimensions,''
Nucl.\ Phys.\ B {\bf 467}, 183 (1996)
[hep-th/9505032].

%
%\bibitem{vanNieuwenhuizen:1996tv}
%P.~van Nieuwenhuizen and A.~Waldron,
%``On Euclidean spinors and Wick rotations,''
%Phys.\ Lett.\ B {\bf 389}, 29 (1996)
%[hep-th/9608174].
%
%
%\bibitem{Cangemi:1992my}
%D.~Cangemi, M.~Leblanc and R.~B.~Mann,
%``Gauge formulation of the spinning black hole in (2+1)-dimensional anti-De Sitter space,''
%Phys.\ Rev.\ D {\bf 48}, 3606 (1993)
%[gr-qc/9211013].






\end{thebibliography}
\end{document}